\documentclass{article}

\def\be{\begin{equation}}
\def\ee{\end{equation}}
\def\bea{\begin{eqnarray}}
\def\eea{\end{eqnarray}}
\def\({\left(}
\def\){\right)}
\def\<{\left<}
\def\>{\right>}

\def\[{\left[}
\def\]{\right]}
\begin{document}

\pagestyle{empty}
\vskip-10pt
\vskip-10pt
\hfill 
\begin{center}
\vskip 3truecm
{\Large\bf
A reparametrization invariant surface ordering
}\\ 
\vskip 2truecm
{\large\bf
Andreas Gustavsson\footnote{a.r.gustavsson@swipnet.se}
}\\
\vskip 1truecm
{\it Institute of Theoretical  Physics,
Chalmers University of Technology, \\
S-412 96 G\"{o}teborg, Sweden}\\
\end{center}
\vskip 2truecm
{\abstract{We introduce a notion of a non-Abelian loop gauge field defined on points in loop space. For this purpose we first find an infinite-dimensional tensor product representation of the Lie algebra which is particularly suited for fields on loop space. We define the non-Abelian Wilson surface as a `time' ordered exponential in terms of this loop gauge field and show that it is reparametrization invariant.}}

\vfill \vskip4pt

\eject
\newpage
\pagestyle{plain}
\section{Introduction}
It has long ago \cite{Nepomechie} been appreciated that, from a geometrical point of view, the natural habitat for rank $2$ tensor gauge fields is the loop space. (The generalization to higher ranks should be straightforward.) But only Abelian tensor gauge fields were considered there. It was later proved \cite{Teitelboim} that tensor gauge fields of rank $>1$ necessarily must be Abelian, as a consequence of the lack of surface ordering that has an invariant significance under reparametrizations of the surface \cite{Teitelboim2}, \cite{Dirac}. 

In this Letter we will present a way to obtain a reparametrization invariant surface ordering, despite the result in \cite{Teitelboim}. Our approach has been inspired by ideas of \cite{Akhmedov} and \cite{Larsson}. These papers were based on ideas from lattice gauge theory \cite{Orland}. However in the references \cite{Akhmedov} interactions could not be seen, which suggests that spacetime fields are not the natural thing to consider. To see the interactions, we should consider fields defined not on points, but on loops, since points can not be ordered on a surface while loops can.

The conventional approach to obtain a surface ordering is to introduce a local one-form connection on space-time, in addition to a local two-form gauge connection, as was done in \cite{Alvarez}. A more recent paper is \cite{Urs}.

In this Letter we introduce a non-Abelian connection one-form directly in loop space, without making any reference whatsoever to local connections in space-time. We have not seen that this loop gauge field has appeared in the literature before, though a corresponding loop space field was discussed in a lattice formulation framework in \cite{Larsson}.

Our ultimate goal is to find a way to define interacting six-dimensional theories with $(2,0)$ supersymmetry \cite{Witten}. This amount of supersymmetry in any non-gravitional theory can be realized on a tensor multiplet which does not contain a space-time one-form gauge field. 

\section{A tensor product representation}\label{repres}
Our main idea is to consider an infinite-dimensional representation of the Lie algebra, and associate the continuous representation indices of the Lie algebra generators with the parameter $s$ say, that parametrizes a corresponding loop $C: s\mapsto C^{\mu}(s)$ in spacetime. We will assume a topologically trivial spacetime. We thus consider an infinite-dimensional tensor product representation 
\bea
t_a(C)=\int ds \lambda_a(s,C)\label{rep}
\eea
where
\bea
[\lambda_a(s,C),\lambda_b(t,C)]=C_{ab}{}^c \delta(s-t)\lambda_c(s,C).\label{comm}
\eea
We notice that $\lambda_a(s,C)$ may be viewed as generators of an infinite-dimensional Lie algebra, provided $C_{ab}{}^c$ are structure constants of a finite-dimensional Lie algebra. 

The generators $t_a(C)$ should not depend on the way we parametrize the loop $C$. Thus if $C(s)=C'(s')$ are two different parametrizations of the same geometrical loop, then we should have $ds \lambda_a(s,C)=ds' \lambda_a(s',C')$. It is easy to see that the algebra is invariant under this reparametrization if one notices that $ds\delta(s-s')$ is invariant. 

We define $\lambda_a(s,C_1)$ for any loop which is homotopic to $C$ by first finding a homotopy map $t\mapsto C_{t}(s)$ such that $C_{t=0}=C$ and $C_{t=1}=C_1$. We then define $\lambda_a(s,C_1)=\lambda_a(s,C)$. The $\lambda_a(s,C_1)$ will then be well-defined only up to an arbitrary reparametrization. But that will not affect the $t_a(C)$, which are reparametrization invariant. We may thus write $t_a(C)=t_a$, and it is easy to see that
\bea
[t_a,t_b]=C_{ab}{}^c t_c \label{lie}
\eea
This explains why we said that Eq. (\ref{rep}) is a representation of the Lie algebra, corresponding to the gauge group. In the particular application to $(2,0)$ supersymmetric theories, these Lie algebras must belong to the $A-D-E$ series \cite{Witten}, \cite{Henningson}.

\section{Gauge covariance}
We will now show what led us to introduce this infinite-dimensional tensor product representation, and how it is to be associated with the non-Abelian loop gauge field. We start by recalling how the gauge symmetry is implemented in Yang-Mills theory, and then generalize this to the rank $2$ tensor gauge field in a natural way. 

In Yang-Mills theory we have the path-ordered exponent
\bea
U(C;y,x)_{ij}=\(P e^{\int_{C} A}\)_{ij}
\eea
associated with a curve $C$ embedded in spacetime, to which we have assigned the indices $i$ and $j$ at its two end-points $x$ and $y$ respectively. We can expand the path ordered exponent as a Dyson series. But what we really have here is an infinite product of matrices 
\bea
U_{ij}(x)=\(e^{A_{\mu}(x)\delta x^{\mu}}\)_{ij}
\eea
which are to be glued together (by the usual matrix multiplication) along a discretized curve that approximates $C$. We then take the limit that the lengths of each line segment tend to zero. The path ordered exponent is thus really an infinite product of operators $U_{ij}$ ordered along a curve $C$.  

The path ordered exponent satisfies the differential equation
\bea
\dot{C}^{\mu}(s)D^A_{\mu}U(C;x,0)=0
\eea
where $C^{\mu}(s)=x$ (and where the dot denotes $d/ds$), and where the covariant derivative is compatible with gauge transformations
\bea
D_{\mu}^{A^g}g=gD^A_{\mu}\label{comp}
\eea
where $g(x)_{ij}=\(e^{\Lambda(x)}\)_{ij}$ is a gauge group element, $\Lambda$ is an element in the corresponding Lie algebra, and $A^g$ is the gauge transformed gauge potential. As a consequence the path ordered exponent transforms as 
\bea
U(y,x)_{ij}\rightarrow U^g(y,x)_{ij}=g(y)_{ii'}U(y,x)_{i'j'}g^{\dag}(x)_{j'j}
\eea
because this transformed object satisfies the gauge transformed differential equation 
\bea
\dot{C}^{\mu}(s)D_{y^\mu}^{A^g} U^g(y,x)=\dot{C}^{\mu}(s)g(y)D_{y^{\mu}}U(y,x)g^{\dag}(x)=0
\eea
as well as the boundary condition
\bea
U^g(x,x)=U(x,x)=1.
\eea

Finally let us examine the compatibility condition (\ref{comp}). Let us make the ansatz
\bea
D_{\mu}=\partial_{\mu}+A_{\mu}.
\eea
Then the compatibility condition reads
\bea
A^g_{\mu} = gA_{\mu}g^{-1}-(\partial_{\mu}g)g^{-1}.
\eea

We now generalize all this to surfaces. Following \cite{Akhmedov} we define a surface ordered exponent (Wilson surface)
\bea
U(S;C_I,C_J,...)_{IJKL...}
\eea
associated with a surface $S$ which we have assumed have a disjoint set of boundary loops $C_I$, $C_J$, ... . To each boundary loop $C_I$, having an orientation that is induced from the orientation of the surface, we associate a continuous index $I$. We would now like to think of this Wilson surface as a path ordered exponent obtained by gluing matrices 
\bea
U_{IJ}(C)
\eea
associated with loops that slices up the surface, this being the natural generalization of the path ordered Wilson loop. 

We postulate the surface exponent to transform under a gauge transformation as 
\bea
U(S,C_I,C_J,...)_{IJ...}\rightarrow g_{II'}(C_I)g_{JJ'}(C_J)...U(S;C_I,C_J,...)_{I'J'...}
\eea
where the gauge transformation is generated by group-elements $g_{IJ}(C)$ defined on closed loops, which should have the property
\bea
g^{\dag}(C)=g(\tilde{C})
\eea
where $\tilde{C}$ is the same loop as $C$ but with its orientation being reversed. A group element should be represented as an infinite tensor product
\bea
g(C)_{IJ}=\bigotimes_{s\in C} g(s,C)_{i_sj_s}
\eea
where each element in the product is of the form 
\bea
g(s,C)_{i_sj_s}=\(e^{\Lambda(s,C)}\)_{i_sj_s}
\eea
and where $s$ parametrizes the loop $C$ as well as elements in the infinite-dimensional multi-index $I=(i_s)$.

The group element can thus be written as the exponent of a Lie algebra element that takes values in an infinite-dimensional tensor representation,
\bea
g(C)_{IJ}=\(e^{\Lambda(C)}\)_{IJ}
\eea
where, schematically,
\bea
\Lambda(C)=\sum_{s\in C} \Lambda^a(s,C) 1\otimes\cdots\otimes t_{a}\otimes \cdots \otimes 1
\eea
where the generator $t_a$ is placed at position $s$. Covariance suggests that the continuum limit of $1\otimes \cdots\otimes t^{a}\otimes \cdots \otimes 1$ should equal $\lambda^a(s,C)$ which was introduced in section \ref{repres}. Thus, in the continuum limit we have
\bea
\Lambda(C)=\int ds \Lambda^a(s,C) \lambda_a(s,C).\label{gauge}
\eea
In order for $\Lambda(C)$ to be reparametrization invariant we must require $\Lambda^a(s,C)$ to be invariant (since $\lambda_a(s,C)ds$ are invariant). We also notice that there is no need to path order the exponent $e^{\Lambda(C)}$ since $[\lambda_a(s,C),\lambda_b(s',C)]=0$ if $s\neq s'$.

The covariant derivative should likewise be defined on loops, and obey a compatibility condition
\bea
D^{A^g}_{\mu}(C)g(C)=g(C)D^A_{\mu}(C).
\eea
Making the ansatz
\bea
D_{\mu}(C)=\int_C ds \(\frac{\delta}{\delta C^{\mu}(s)} + A_{\mu}(s,C)\)
\eea
we find from the compatibility condition that
\bea
A^g_{\mu}(C)=g(C)A_{\mu}(C)g^{-1}(C)+g(C)\partial_{\mu}(C)g(C)^{-1}.
\eea
where we have introduced 
\bea
\partial_{\mu}(C)\equiv \int ds\frac{\delta}{\delta C^{\mu}(s)}
\eea
and
\bea
A_{\mu}(s,C)&=&A_{\mu}^a(s,C)\lambda_a(s,C)\cr
A_{\mu}(C)&=&\int ds A_{\mu}(s,C)
\eea
is a gauge potential defined on loops rather than at points, in the same fashion as the gauge parameter $\Lambda(C)$ was defined in Eq. (\ref{gauge}). Just like the group elements $g(C)$ should not not depend on the parametrization of the loop, the gauge potential $A_{\mu}(C)$ should not either. Thus if $C(s)=C'(s')$ are two different parametrizations, then we should have $ds A_{\mu}(s,C)=ds' A_{\mu}(s',C')$.

\section{Surface ordering}
We now define the Wilson surface (or surface holonomy or surface ordered exponent) as 
\bea
U(S;C_1,C_2,...)=P_t \exp \int dt\int ds\frac{\partial C^{\mu}_t(s)}{\partial t}A^a_{\mu}(s,C_t)\lambda_a(s)\label{Wilson}
\eea
where $P_t$ orders with respect to the parameter $t$. In the Abelian case the ordering operator is of course superfluous, and if we introduce the constraint $A_{\mu}(s,C)=B_{\mu\nu}(C(s))\dot{C}^{\nu}(s)$ we see that our definition reduces to the usual Abelian Wilson surface for what will turn out to be the usual local two-form gauge field $B_{\mu\nu}$,
\bea
U(S;C_1,C_2,...)=\exp \int dt\int ds \frac{\partial C^{\nu}_t}{\partial t}\frac{\partial C^{\mu}_t}{\partial s}B_{\mu\nu}(C_t(s)).
\eea
which clearly is reparametrization invariant.\footnote{In the Abelian case $ds A_{\mu}(s,C)=ds B_{\mu\nu}(C(s))\dot{C}^{\nu}(s)$ is reparametrization invariant. Thinking of the Abelian Lie algebra as being generated by the reparametrization invariant element $1$, it seems natural to try to generalize this situation to the non-Abelian by taking invariant generators $t^a(s,C)=\lambda^a(s,C)ds/d\bar{s}$ where $\bar{s}$ is some fixed but arbitrary parametrization, and assuming that $ds A^a_{\mu}(s,C)$ is invariant, to make $A_{\mu}(C)$ invariant under reparametrizations. But it seems strange that the Wilson surface would depend on such an arbitrary parametrization $\bar{s}$. We could try to fix this arbitrary parametrization by taking it to be the proper distance along the loop. But that just makes things worse. It would make the generators, and consequently the Wilson surface, metric dependent, which we certainly should not have as the Wilson surface is a topological quantity. I would like to thank Urs Schreiber for making me aware of these things.}

Now let us turn to the non-Abelian case, and let us think of the parameter $t$ as time. The philosphy to show reparametrization invariance is now to show that the Wilson surface does not depend on the particular way we happen to choose the time slicing. That is, we show that we may deform the constant time loops $C_t$. We will refer to this property as `path independence', using the terminology of \cite{Teitelboim}. Here we mean by a loop the geometrical loop plus its parametrization. Thus a deformation of a loop need not necessarily involve a geometrical deformation -- it may also be a reparametrization of the loop. 

Viewing $t$ as time, the path ordered exponent is precisely the time evolution operator, which satisfies the Schroedinger equation,\footnote{The surface holonomy can be seen as a Wilson line in loop space with metric $A_{\mu}(C)\cdot B^{\mu}(C)=\int ds g_{\mu\nu}(C(s))A^{\mu}(s,C)B^{\nu}(s,C)$, where $g$ is the space-time metric. We thus write the surface holonomy as 
\bea
U(S;C_1,C_2,...)=P_t \exp \int dt \frac{dC^{\mu}_t}{dt}\cdot A_{\mu}(C_t).
\eea
The surface holonomy obeys
\bea
\frac{d{C_t}^{\mu}}{dt}\cdot D_{\mu}(C_t) U(S;C_t,...)=0
\eea
i.e. it is covariantly constant along a loop in loop space (which means a surface in space-time). This condition can be recast in the form of a Schroedinger equation.} 
\bea
\frac{dU(t)}{dt}=H(t) U(t)  
\eea
with Hamiltonian
\bea
H(t)=\int ds \frac{\partial C^{\mu}_t(s)}{\partial t}A_{\mu}(s,C_t)
\eea
Here $U(t):=U(S;C_t,...)$. Identifying this Hamiltonian with
\bea
H(t)=\int ds \(\frac{\partial C^{\perp}_t(s)}{\partial t} {\cal H}_{\perp}+\frac{\partial C^s_t(s)}{\partial t} {\cal H}_s\)
\eea
where
\bea 
\delta C^{\perp}&\equiv&\delta C^{\mu} n_{\mu}\cr
\delta C^s&\equiv &\delta C^{\mu}\partial^s C_{\mu}(s)
\eea
with $n^{\mu}=n^{\mu}(s)$ denoting the unit normal vector to the time slice lying in the surface $S$; $n^{\mu}\partial^s C_{\mu}(s)=0$, $n^{\mu}n_{\mu}=1$, one finds
\bea
{\cal H}_{\perp}&=&n^{\mu}A_{\mu}\cr
{\cal H}_s&=& 0.
\eea
The condition of 'path independence' in the sense that different time slicings yield the same result, amounts in this special situation to the single condition \cite{Teitelboim},
\bea
[{\cal H}_{\perp}(t,s),{\cal H}_{\perp}(t,s')]=0.
\eea
which obviously is satisfied by
\bea 
{\cal H}_{\perp}(s,t)&=&n^{\mu}A^a_{\mu}(s,C_t)\lambda_a(s,C_t)
\eea
if and only if 
\bea
[\lambda_a(s,C_t),\lambda_b(s',C_t)]=0{\mbox{ whenever }}s\neq s' \label{comm1}
\eea
If on the other hand $s=s'$, then we have a commutator between two quantities ${\cal H}_{\perp}(t,s)$ that are equal. So the commatator vanishes identically with this construction. 

One should also notice that this condition of path independence would not be satisfied if we would take
\bea
{\cal H}_{\perp}(s,t)&=&n^{\mu}A^a_{\mu}(s,C_t)t_a
\eea
with $t_a$ being Lie algebra generators, unless that Lie group is Abelian \cite{Teitelboim}.

\section{A dynamical theory in loop space}
We now proceed to find the gauge field strength. There are several ways to obtain this. The easiest way is probably to compute the commutator 
\bea
[D_{\mu}(C),D_{\nu}(C)]=F_{\mu\nu}(C).
\eea
Another way is to compute the variation of the Wilson surface and define the loop gauge field strength by
\bea
\frac{\delta U(S)}{\delta C^{\mu}_t(s)}=\int ds' U(0,t)F_{\mu s,\nu s'}(C_t)\frac{\partial{C_t}^{\nu}(s')}{\partial t}U(t,1)
\eea
where obviously $U(0,t)$ denotes the Wilson surface ordered in time where we integrate it up to time $t$ (and similarly for $U(t,1)$, so that in particular $U(0,t)U(t,1)=U(S)$.) Either way we get the non-Abelian loop field strength as
\bea
F_{\mu\nu}(C)=\partial_{\mu}(C)A_{\nu}(C)-\partial_{\nu}(C)A_{\mu}(C)+[A_{\mu}(C),A_{\nu}(C)]
\eea
which we may write as
\bea
F_{\mu\nu}(C)=\int ds \int dt F_{\mu s, \nu t}(C)
\eea
where
\bea
F_{\mu s,\nu t}(C)=\frac{\delta A_{\nu}(t,C)}{\delta C^{\mu}(s)}-\frac{\delta A_{\mu}(s,C)}{\delta C^{\nu}(t)} + [A_{\mu}(s,C),A_{\nu}(t,C)].
\eea

The equation of motion must be gauge covariant. If we require it to involve at most two derivatives there is only one such equation that we can write down, namely\footnote{Here we take the point of view that Eq. (\ref{comm}) defines an infinite-dimensional Lie algebra. The adjoint representation of this Lie algebra in which the gauge field tke its values is given by the commutator $[t_a(s,C),\cdot]$.}
\bea
D_{\mu}(C)F^{\mu\nu}(C)=0
\eea
Spelling it out, we have
\bea
\partial_{\mu}(C)F^{\mu\nu}(C)+[A_{\mu}(C),F^{\mu\nu}(C)]=0.\label{eom}
\eea

It might be possible to realize $(2,0)$ supersymmetry on non-Abelian loop fields in loop space. In the Abelian case this is quite obviously the case. Here the (space-time) tensor multiplet fields (on which $(2,0)$ SUSY is realized) are a self-dual two-form gauge connection $B_{\mu\nu}(x)$, five scalar fields $\phi^A(x)$ and a symplectic Majorana-Weyl spinor $\psi(x)$. Linear self-duality is implemented on the field strength $H=dB$ as $H_{\mu\nu\rho}(x)=\frac{1}{6}\epsilon_{\mu\nu\rho\kappa\tau\sigma}H^{\kappa\tau\sigma}(x)$. We can define corresponding Abelian tensor multiplet loop fields as
\bea
A_{\mu}(C)&=&\int ds \dot{C}^{\nu}(s)B_{\mu\nu}(C(s))\cr
\phi^{\mu}(C)&=&\int ds \dot{C}^{\mu}(s)\phi(C(s))\cr
\psi^{\mu}(C)&=&\int ds \dot{C}^{\mu}(s)\psi(C(s))
\eea
in terms of which the self-duality constraint becomes $F_{\mu\nu}(C)=\frac{1}{6}\epsilon_{\mu\nu\rho\kappa\tau\sigma}F^{\rho\kappa\tau\sigma}(C)$, where we define
\bea
F_{\mu\nu}(C)&=&\int ds\int dt F_{\mu s,\nu t}(C)\cr 
&=&\int ds H_{\mu\nu\rho}(C(s))\dot{C}^{\rho}(s)\cr
F_{\mu\nu\rho\sigma}(C)&=&\int ds H_{[\mu\nu\rho}(C(s))\dot{C}_{\sigma]}(s).
\eea
It should be noticed that these loop fields are in one-to-one correspondence with the local fields. From any of these Abelian loop fields, we can get back the corresponding local field by computing a functional integral. For instance, for the scalar field, we have
\bea
\phi(x)g^{\mu\nu}(x)\sim\lim_{\epsilon\rightarrow 0}\int_{L(C)\leq\epsilon} D_x C \phi^{\mu}(C){C}^{\nu}(s)
\eea
Here $g_{\mu\nu}(x)$ is the space-time metric tensor, and the functional integral is over all loops $C$ of some length scale $L(C)$ less than $\epsilon$, and which are `close to' $x$ in a suitable sense \cite{Nepomechie} (for instance those which goes through $x$). To see this, one uses \cite{Nepomechie}
\bea
\int D_x C \dot{C}^{\mu}(s)\dot{C}^{\nu}(s')\sim \delta(s-s')g^{\mu\nu}(x).
\eea

Once supersymmetry has been reformulated in terms of these Abelian loop fields, the generalization to the non-Abelian case should be obtained by replacing the Abelian loop fields with non-Abelian loop fields, and ordinary loop derivatives with covariant loop derivatives. To check the closure, one also has to figure out how to implement the self-duality constraint on the non-Abelian loop gauge field. We will return to this point in a future paper. 

As guidance one also has that the non-Abelian supersymmetry variations should be such that, upon dimensional reduction in one dimension, one gets the supersymmetry variations of $d=5$ super Yang-Mills with $16$ supercharges. Under such a dimensional reduction the loop space fields and the gauge covariant derivative reduce to ordinary local space-time fields and gauge covariant derivative. Compactifying the $x^5$ dimension on a circle and taking the loops to be straight lines in the $x^5$-direction, and defining (for $m=0,1,...,4$)
\bea
A_{m}(x)=\int ds A^a_m(s,C_x)\lambda_a(s,C_x)
\eea
where $C_x$ denotes the loop along the $x^5$ axis whose transverse coordinates are $x$, we find in the compactification limit where the Kaluza-Klein modes become infinitely heavy, that $A_m(x)=A_m^a(x) t_a$ will be the gauge field in five dimensional Yang-Mills theory. Notice that for a straight loop (a loop obtained by identifying two end-points of a straight line) we need just associate one index. So the matrices $t_a$ may be in the fundamental representation of the gauge group, or in any other representation. Thus they need not be in the infinite-dimensional tensor product representation that we would get from a smoothly curved loop.

\end{document}